# Tuning Bandgap and Energy Stability of Organic-Inorganic Halide Perovskites through Surface Engineering

Rahul Singh,[1] Prashant Singh,[2] and Ganesh Balasubramanian[3]
[1]Department of Mechanical Engineering, Iowa State University, Ames, IA 50011
[2]Division of Materials Science & Engineering, Ames Laboratory, Ames, Iowa 50011, USA
[3]Department of Mechanical Engineering & Mechanics, Lehigh University, Bethlehem, PA 18015

**Abstract**

Organohalide perovskite with a variety of surface structures and morphologies have shown promising potential owing to the choice of the type of heterostructure dependent stability. We systematically investigate and discuss the impact of 2-dimensional molybdenum-disulphide ($MoS_2$), molybdenum-diselenide ($MoSe_2$), tungsten-disulphide ($WS_2$), tungsten-diselenide ($WSe_2$), boron-nitiride (*BN*) and graphene monolayers on band-gap and energy stability of organic-inorganic halide perovskites. We found that $MAPbI_3ML$ deposited on *BN-ML* shows room temperature stability (-25 meV~300K) with an optimal bandgap of ~1.6 eV. The calculated absorption coefficient also lies in the visible-light range with a maximum of 4.9 x $10^4$ $cm^{-1}$ achieved at 2.8 eV photon energy. On the basis of our calculations, we suggest that the encapsulation of an organic-inorganic halide perovskite monolayers by semiconducting monolayers potentially provides greater flexibility for tuning the energy stability and the bandgap.

**TOC Graphic:**

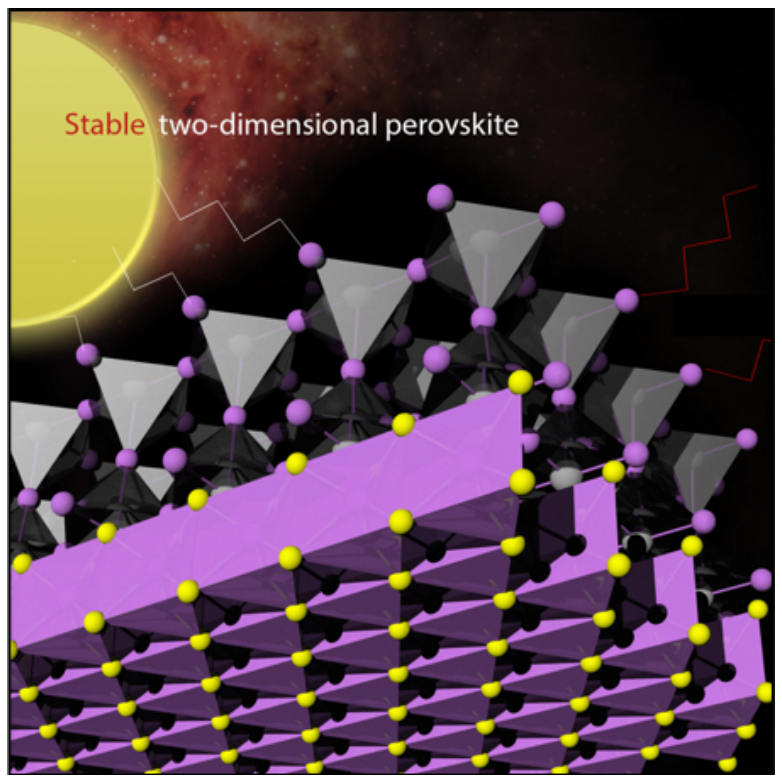

In the past few years, the solar cell community has witnessed an exceptional emergence of a new family of solar cell materials[1–3] namely organic-inorganic halide perovskite based solar cells. Metal-halide perovskites, with the formula $ABX_3$ (e.g., $A$=methylammonium ($MA$), $B = Pb$, and $X = I$) are rapidly gaining attention for thin film photovoltaics due to their long carrier diffusion lengths,[4] defect tolerance enabling versatile solution or vapor processing,[5,6] and strong optical absorption.[7] Within just four years, the conversion efficiency has increased dramatically to 20.1%.[8] This development is believed to be a result of a unique supportive combination of different properties, including the favorable balance between strong absorption and long carrier lifetimes,[9] the efficient-transport,[10–12] and the fault tolerance[13] of these materials. Recently, perovskite-silicon tandems have achieved record efficiencies of 23.6% and 26.4% for monolithically integrated[14] and mechanically stacked configurations.[15] Additionally, recent developments in the stability and efficiency of low bandgap tin-based perovskites have also drawn attention for making highly efficient and potentially low-cost perovskite–perovskite tandems.[16,17]

Currently, energetic instability of $MAPbI_3$ is well known, which arises both from extrinsic and intrinsic causes. Extrinsically, $MAPbI_3$ is sensitive to moisture, UV exposure, and oxygen.[3,18] Another important aspect that contributes significantly to the instability is the moderate crystal quality,[19] that ignites consequently an additional mixture of instability issues. Here, we study the energy stability and electronic structure of monolayer (ML) $MAPbI_3$ deposited on 2-dimensional molybdenum-disulphide ($MoS_2$), molybdenum-diselenide ($MoSe_2$), tungsten-disulphide ($WS_2$), tungsten-diselenide ($WSe_2$), boron-nitiride ($BN$) and graphene monolayers. This study will be an important contribution in developing stable organic-inorganic halide perovskite variants with desired properties for optoelectronic applications.

We use density functional theory (DFT)[20,21] based Vienna ab initio simulation package (VASP)[22–24] to examine the stability, electronic-structure and optical/transport behavior of $MAPbI_3$*ML based* heterostructures. We construct [001] terminated 2D OIHPs from bulk $MAPbI_3$ and deposited *on* X-ML (=$MoS_2$, $MoSe_2$, $WS_2$, $WSe_2$, *BN*, and *Graphene)* surfaces to prepare X-ML/ $MAPbI_3$*ML heterostructure*. The $MAPbI_3$*ML deposited* ($MoS_2$, $MoSe_2$, $WS_2$, $WSe_2$), *BN* and *Graphene* supercells have 174, 213, and 339 atoms, respectively. For geometry optimization and electronic structure calculations, we use the projected augmented-wave (PAW) basis[25] and the Perdew–Burke–Ernzerhof (PBE)[22] exchange-correlation functional. The charge and forces are converged to $10^{-5}$ eV and 0.01eV/Å, respectively, using energy cut-off of 800 eV. The Monkhorst-Pack[25] k-mesh grid of 5×5×3 is used for all the structures. The tetrahedron method with Blöchl corrections is used to calculate the density of states (DOS).[22] The optical properties in terms of absorption spectra are calculated by obtaining the frequency dependent dielectric matrix using VASP.[26] The Brillouin-zone (BZ) is sampled with a Gamma-centered $k$-mesh with 70 $k$ points. For the transport calculations a non-shifted mesh with 70000 $k$ points is used. The necessary derivatives were then calculated on an FFT grid. The thermoelectric properties are calculated using the BoltzTrap[27] code interfaced with VASP.

The most widely used methylammonium cation (*MA*; $CH_3NH_3^+$; $R_{\mathrm{MA}} = 0.18$ nm), has achieved efficiencies up to 20.1% and above for $MAPbI_3$ devices.[28,29] The cubic $MAPbI_3$ yields an optimal optimal bandgap of 1.1-1.4 eV dictated by the Shockley-Queisser limit for a single junction solar cell.[28,29] Clearly, if we can stabilize the $MAPbI_3$ in the ambient conditions (at room temperature), one may speculate enhanced light harvesting across the spectrum. To that end, we employ first-principles density functional theory to examine stability, electronic-structure and transport properties of *X-ML/*$MAPbI_3$ML (X= $MoS_2$, $MoSe_2$, $WS_2$, $WSe_2$, *BN*, and *Graphene)* based heterostructures, shown in Figure 1.

The binding energy is a very important criteria to predict that if any chemical structure can be chemically and energetically stable or not. The binding energy of six heterostructures is calculated as:

$$E_b = E_{Total}(X - ML/MAPbI_3ML) - E(MAPbI_3ML) - E(X - ML)$$

here, X=$MoS_2$, $MoSe_2$, $WS_2$, $WSe_2$, *BN*, and *Graphene*. The $E_b$ in the expression above is the binding energy. We found that heterostructures are more stable when we chose $PbI_2$/*X-ML* slabs than the corresponding *MAI/X-ML* slabs. The reason should be that the interaction between *MA* ions of the *MAI-X-ML* with other ions is through either van der Waals (vdW) or hydrogen bond, which is rather weak. The interaction between the *Pb* atom in $PbI_2$/*X-ML* slabs is through chemical bonding, which connects *X-ML* and perovskite as a bridge. The binding energy is shown in Figure 2 (green circles). Of the six structures, $MAPbI_3$*ML/Graphene* shows positive binding energy of +0.2 meV, which indicates the energy instability towards formation of the heterostructure, i.e.,

*$MAPbI_3ML$/Graphene* is not favorable. However, other five cases, viz., *$WS_2$*, *$WSe_2$*, *$MoS_2$*, *$MoSe_2$* and *BN* cases show negative binding energies, i.e., favorable to the formation of the heterostructure. The *$WS_2$*, *$WSe_2$*, *$MoS_2$*, and *$MoSe_2$* are almost energetically degenerate. On the other hand, the binding energy of the *BN-ML/$MAPbI_3ML$* is -25 meV, which is equivalent to 300 K on the temperature scale. This shows that *BN-ML/$MAPbI_3ML$* can be stabilized at room temperature. We also analyze the binding energies with spin-orbital coupling, and we observe very small deviation of binding energies, e.g. for *BN-ML/$MAPbI_3ML$* $E_b \sim 23.5$ meV.

The bandgap is another important characteristic property of photovoltaic materials. The band-gaps of the six structures, in Figure 2, is shown by red squares. The heterostructure *Graphene/$MAPbI_3$* has a bandgap of 0.18 eV. In the presence of *$MAPbI_3ML$*, the bandgap of graphene opens due to the stronger interlayer interaction. Similar results have been reported before: a bandgap of 0.1 eV[30] when graphene lies on *BN-ML* substrate, and similarly, a bandgap of 0.25 eV[31] is observed for silicon carbide substrate. *Graphene/$MAPbI_3ML$* can be used to create semiconducting graphene materials that maintain the exceptional transport property. As we move to different structures, the bandgap is higher. *$MoS_2ML/MAPbI_3ML$* has a bandgap of 1.40 eV, which is smaller than the reported bandgap of $MoS_2$ (1.8 eV[32]) and higher than 2D *$MAPbI_3$* (1.34 eV[33]). So, the bandgap is closer to the bandgap of 2D *$MAPbI_3$*. Similar observation is obtained for *$WSe_2ML/MAPbI_3ML$*, where in the bandgap is 1.40 eV. Theoretical calculation shows that the solar efficiency for a single bandgap semiconductor, is maximum 33% at a bandgap 1.4 eV.  So, for solar applications, *$MoS_2ML/MAPbI_3ML$* and *$WSe_2ML/MAPbI_3ML$*, provide an alternative approach to make new novel materials. However, for *$WS_2ML/MAPbI_3ML$* and *$MoeS_2ML/MAPbI_3ML$*, there are some hybridization effects that reduce the bandgap from the original 2D structure band gaps. The bandgap in *$WS_2ML/MAPbI_3ML$* and *$MoSe_2ML/MAPbI_3ML$* is 0.88 eV and 1.05 eV, respectively. These offer low possibilities in solar cell applications. Among the six heterostructures, the *BN-ML/$MAPbI_3ML$* shows room temperature stability with a bandgap of 1.68 eV, which is very close to the optimum bandgap of 1.40 eV[34] for photovoltaic applications. Since *BN-ML/$MAPbI_3ML$* is the only case with large energy stability and optimal bandgap, so here we analyze and discuss the structural, electronic, optical and transport properties of *BN-ML/$MAPbI_3ML$*.

We constructed the *BN-ML/$MAPbI_3$ML* supercell using the experimental *$MAPbI_3$* and *BN* structures. We kept unit cell periodic in x-y plane and vacuum of at least 18 Angstrom in z-direction. We perform full (lattice + ionic) optimization of *BN-ML/ $MAPbI_3$ML* to calculate equilibrium ground-state structure and properties. In Figure 3, structural parameters, bond-length and bond-angels of *BN-ML/$MAPbI_3$ML* are shown. The *BN-ML/$MAPbI_3$ML* supercell consists of 213 atoms (C=7; N=7; H=42; Pb=14; I=35; Mo=28; BN=108 atoms). We kept unit cell periodic in x-y plane and vacuum of at least 23.5 Angstrom in z-direction. We also make the shell large enough so that lattice mismatch between *BN-ML* (2 atom per cell) and *$MAPbI_3$* (12 atom per cell) reduces to a value less than 1%. We further relax the supercell to remove any strain left during unit cell construction due to lattice mismatch. The (a×b×c) dimensions of the cell before and after relaxation are (9.03965Å x 34.86520Å x 23.49685Å) and (9.05332 Å x 35.12486 Å x 26.00071 Å), respectively, also given in Figure 2. We observe very small cell expansion in a and b lattice parameters, which changes by $[(a_{relax} - a_{unrelax})/ a_{relax}]\% = 0.1$; $[(b_{relax} - b_{unrelax})/ b_{relax}]\% = 0.7$, respectively.

We observe slightly increased volume of the relaxed structure compared to unrelaxed one. The major changes in structural behavior occurs at the interface boundary of *$MAPbI_3$ML* and *BN-ML*, where bond lengths of *$MAPbI_3$ML* show significant distortion. For example, the bond lengths of *Pb-I* bond that are closer to BN are in the range of 3.2 – 3.25 Å, while the bond lengths of Pb-I bonds that are facing BN layer are in the range of 3.11-3.35 Å. This distortion is indicative of increased interlayer chemical interactions. The interfacial *Pb* atoms obviously move towards the perovskite, and the corresponding bond lengths of *Pb-B/Pb-N* and *I-B/N* are 3.023/3.004 Å and 2.963/3.041 Å, respectively. The relative shorter interfacial bond lengths indicate a strong interaction between the two surfaces. In the interfacial region, the surface $PbI_6$ of perovskite is slightly distorted, and the opposite (*I-Pb-I; Pb-I-Pb)/I-Pb-N* bond angle sharply decrease to 170.529/164.49°, and (B-N-B; N-B-N) changes to (120.0; 119.01°) from 120°, respectively. In Figure 2 (b) We also observe special arrangement of the *$MAPbI_3$ML* on *BN-ML*. The *$MAPbI_3$ML* shows a maze type arrangement, where *Pb* on every alternate *$MAPbI_3$* tetrahedron (T) is bonded to N and forms an octahedron (O). The average bond-length of octahedrally bonded *Pb* is 3.218 Å, whereas tetrahedrally bonded *Pb* shows average bond-length of 3.245 Å.

The band structure, density of states, charge density and absorption coefficient for *BN-ML/MAPbI$_3$ML* are shown in Figure 4. In Figure 4(a), the band-structure and density of states for *BN-ML/MAPbI$_3$ML* shows a direct bandgap of 1.68 eV along *Γ*-point of the Brillouin zone. The right panel in Fig. 4(a) shows the projected density of states for *I, Pb* and *BN*. Among them, *Pb-s* states just below the Fermi-level are chemically most active and show weak hybridization. Other states include contribution from p-states represented by blue color, red color for d-states, and green for s-states. The majority of contribution is from p-states as seen from the Figure 3(a) for all the three cases. *I-p* and *Pb-p* have the maximum states near the fermi level with *I-p* states occupying valence band and *Pb-p* occupying conduction band. This behavior has been observed before both in 3D MAPbI$_3$ and 2D MAPbI$_3$. For BN, projected density of states shows the contribution of BN above +4.0 eV and below -0.4 eV. The blue color in the density of states show the major contribution from its *p*-states. Although the BN are very localized near valence band, Valance band maximum (VBM) are mostly dominated by *I-p* states whereas *Pb-p* states dominate Conduction band minimum (CBM). An analysis of the density of states (DOS) provides a better understanding of the bandgap variations. It depicts an additional peak for states within *MAPbI$_3$* extending from the main VBM peak. This feature is largely attributed to the *I-5p* states, however, antibonding interactions of *Pb-6s* orbitals also have significant contributions. Likewise, the *Pb-6s* orbitals are seen at a much lower energy level and thus are less inclined to interact with *I-5p* orbitals and constitute the rapid increase of the VBM of *MAPbI$_3$*. *Pb* and *I* are heavy atoms, such that the *6p* shells in Pb stabilize the *Pb-6s* orbital rendering it less likely to react or share its electrons (*6s* inert pair effect). This results in an increased the bandgap as the VBM shift is approximately 0.5 eV greater than the CBM shift.

Figure 4(b) shows the charge distribution of the *BN-ML/MAPbI$_3$ML* system and portrays an electronic charge density difference for *BN-ML/MAPbI$_3$ML*, highlighting the weak interaction and negligible overlap of electron orbitals between the organic component and inorganic *Pb-I* octahedra. The charge density shown here is the charge density difference of *BN-ML/MAPbI$_3$ML* with respect to *BN-ML* and *MAPbI$_3$ML* charge densities. The distorted green blobs near the interface shows charge density that altered due to interlayer interaction with *BN-ML*. The pink blobs in charge density comes from *BN-sp* states, which is the result of charge-transfer from *MAPbI$_3$ML* and *BNML*. The *MA* cation has shown no direct contribution toward electronic properties.

Hybrid perovskites exhibit strong optical absorbance, allowing for a much-reduced thickness necessary to efficiently facilitate collection of charge carrier. It also gives information regarding the energy efficiency of the material for solar cell applications. The absorption coefficient is useful when the value is greater than zero in the visible spectrum of sunlight. Figure 4(c) shows the comparison of the optical absorption coefficient of the three 2D layers namely the *BN* layer, 2D *$MAPbI_3$* and *BN-ML/$MAPbI_3$ML*. It is a measure of the penetration of light at specific wavelength (energy) before it gets absorbed. The optical absorption increases exponentially in the below bandgap region, with no presence of optically detected (deep) states. The absence of detectable sub-bandgap absorption below photon energies of 1.5 eV for perovskites makes it a promising candidate for tandem devices, where energy band alignments of commonly employed hybrid perovskite materials. Both *$MAPbI_3$* and *BN-ML/$MAPbI_3$ML* show positive absorption coefficients in the visible spectrum and hence are useful for solar cell applications. Although the value is slightly smaller for *BN-ML/$MAPbI_3$ML* (maximum = 4.9 x $10^4$ $cm^{-1}$ at 2.8 eV photon energy) compared to *$MAPbI_3$* (maximum = 5.2 x $10^4$ $cm^{-1}$ at 2.8 eV photon energy), the absorption coefficient is broader compared to *$MAPbI_3$*. The absorption peak for *BN-ML/$MAPbI_3$ML* is sharp, indicating a direct bandgap as shown for *$MAPbI_3$ML* too.[35–37] Absorption spectra for *BN-ML/$MAPbI_3$ML* is nearly identical to *$MAPbI_3$*,[36,38] where *$MAPbI_3$* has a sharp absorption onset with an Urbach energy of 15 meV.[39] The *$MAPbI_3$*-based devices typically achieve absorption up to the tail end of the red region of the spectrum, approximately 800 nm.[39] In such cases, larger shift in the VBM is observed than that of the CBM, effectively reducing the band gap.

The electronic transport properties are dependent on the effective mass obtained from the band structure. The effective hole and electron mass of *BN-ML/$MAPbI_3$ML* in (x, y) plane at X- and M-point of Brillouin zone are [(0.92;0.076)$m_h^*$; (0.043; 0.25) $m_e^*$] *and [*(0.384;0.145)$m_h^*$; (0.156; 0.042) $m_e^*$], respectively and are shown in Table. 1. We observe an anisotropic nature of the effective mass in the two planar directions, which can be used to tune directional transport in *BN-ML/$MAPbI_3$ML*. The values are higher in *x* direction compared to *y* direction. As effective mass directly affects the mobility of the electrons/holes, we can infer that mobility in the *y*-direction will be larger relative to the *x*-direction. This feature is reflected by the calculated effective masses seen for various symmetries across the band structure. Thus, it is clear that *BN-ML/$MAPbI_3$ML* ought to provide improved hole transport and accordingly reduced electron transport.

In TE materials, a high Seebeck coefficient (*S*) at a given carrier concentration results from a high overall DOS effective mass ($m_d$*). However, $\sigma$ decreases with increasing $m_d$*, and also depends on the inertial effective mass, *m**. Ioffe showed empirically that for doped semiconductors to be good TEs, the sweet-spot for carrier concentration, $n \sim 10^{18}$–$10^{20}$ per $cm^3$, corresponding to degenerate semiconductors or semimetals.[40] Here, we predict a similar range for $n_{hole}$ and $n_{electron}$. As the doping concentration increases, $\sigma$ increases and *S* decreases. The contribution to the overall Seebeck coefficient depends on both the positive and negative charge carriers. At low temperatures the population of minority carriers is small, meaning that they will not contribute much to the overall S. At higher temperatures, though, a broadening Fermi distribution leads to an exponential increase in minority carrier conductivity resulting in a reduction (and therefore peak) in the S. Typical degenerate thermoelectric semiconductors display magnitude of the Seebeck coefficient, which rises linearly with temperature to a maximum followed by a decrease.

In Figure 5, we present the transport properties in *BN-ML/MAPbI$_3$ML* heterostructure in the form of Power factor. Power factor ($P = S^2\sigma/\tau$, where S is Seebeck coefficient, σ is electrical conductivity and τ is relaxation time) as a function of charge carriers and temperature is reported to focus on the thermoelectric properties of *BN-ML/MAPbI$_3$ML.* In Figure 5(a) the value of P increases with increase in the value of temperature. As the value of temperature increases, the intrinsic carrier concentration may become comparable to the dopant concentration, and significant electrical conduction could occur by both electrons and holes. Figure 5(b) shows the variation of P with carrier concentration (*n*). As the intrinsic carrier concentration increases, we see an increasing trend in the value of P and then after reaching a limit, it starts decreasing. This can happen since the Seebeck coefficient for the electrons is negative, the average S is decreased, which affects the value of Power factor and it decreases after a certain value of charge carrier.

We systematically investigate the energy stability and electronic-structure of various organic-inorganic halide perovskite based heterostructures. The *BN-ML/MAPbI$_3$ML* is the most stable among six heterostructures for which we perform comprehensive structural, electronic-structure and transport analysis. We notice that *MAPbI$_3$ML* shows very unique structural arrangement on *BN-ML,* where *Pb*-atoms at *BN*-interface bonds to specific N-atoms forming periodically arranged *Pb(I$_5$N)* octahedron. This unique chemical bonding leads to large energy-stability of *BN-ML/MAPbI$_3$ML* compared to others. The calculated bandgap and absorption coefficients lie in the optimum performance range, which is important for the targeted optoelectronic applications. We believe that the results are important in understanding the relationship among various heterostructures of organohalide lead perovskites. This work also provides a fundamental background for the development of organohalide lead perovskite with enhanced performance and improved stability.

**Acknowledgment**

The research was supported, in part, by the National Science Foundation (NSF) grant no. CMMI-1404938. The work at Ames Laboratory was supported by the U.S. Department of Energy (DOE), Office of Science, Basic Energy Sciences, Materials Science & Engineering Division, which is operated by Iowa State University for the U.S. DOE under contract DE-AC02-07CH11358.

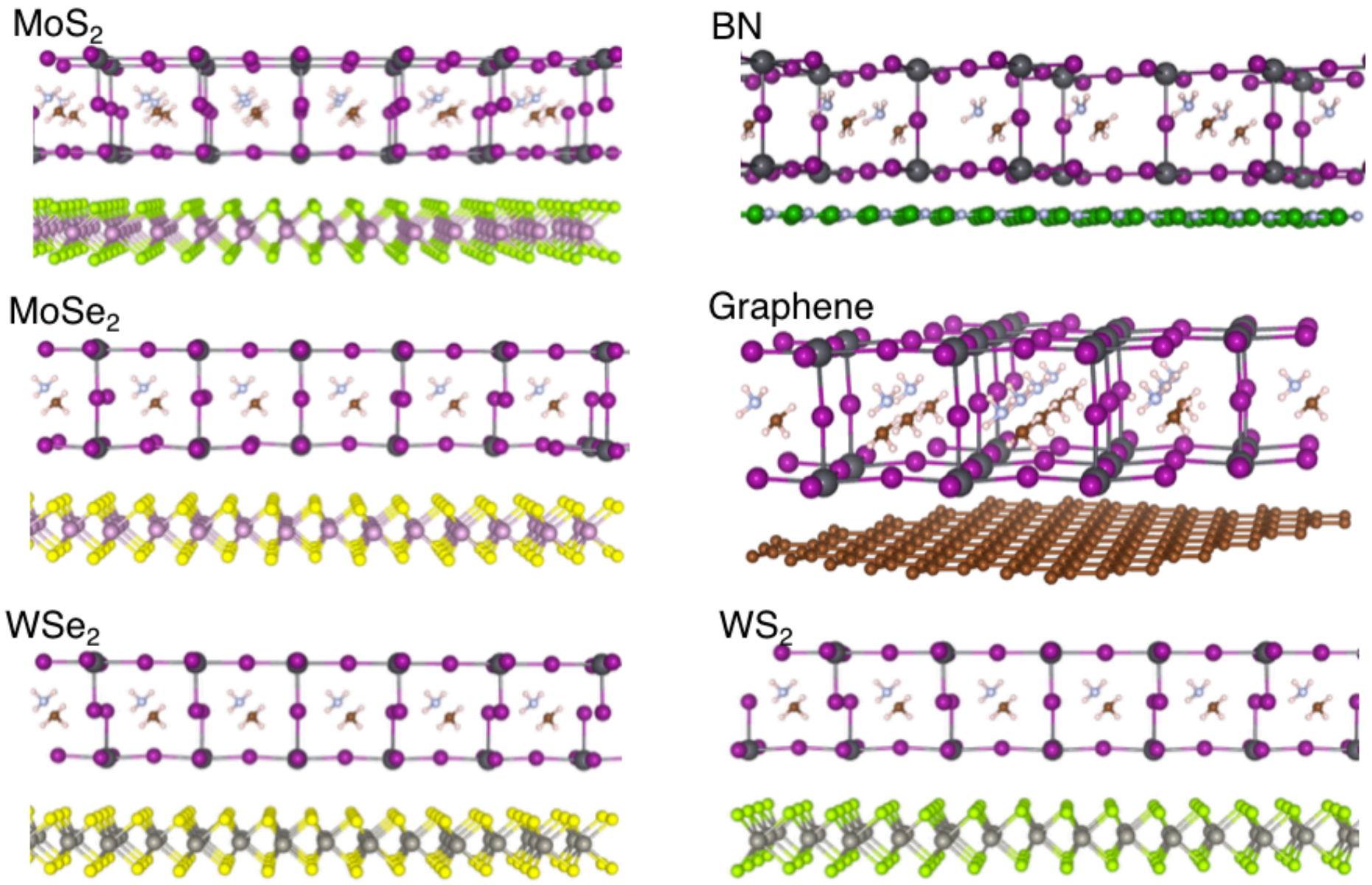

**Figure 1:** Schematic diagram of the six layered structures chosen in this study. A monolayer of two-dimensional $MAPbI_3$ is deposited on different two-dimensional (2D) materials. The 2D materials include molybdenum-disulphide ($MoS_2$), boron-nitride (BN), molybdenum-disulphide ($MoSe_2$), graphene, tungsten-diselenide ($WSe_2$) and tungsten-disulphide ($WS_2$).

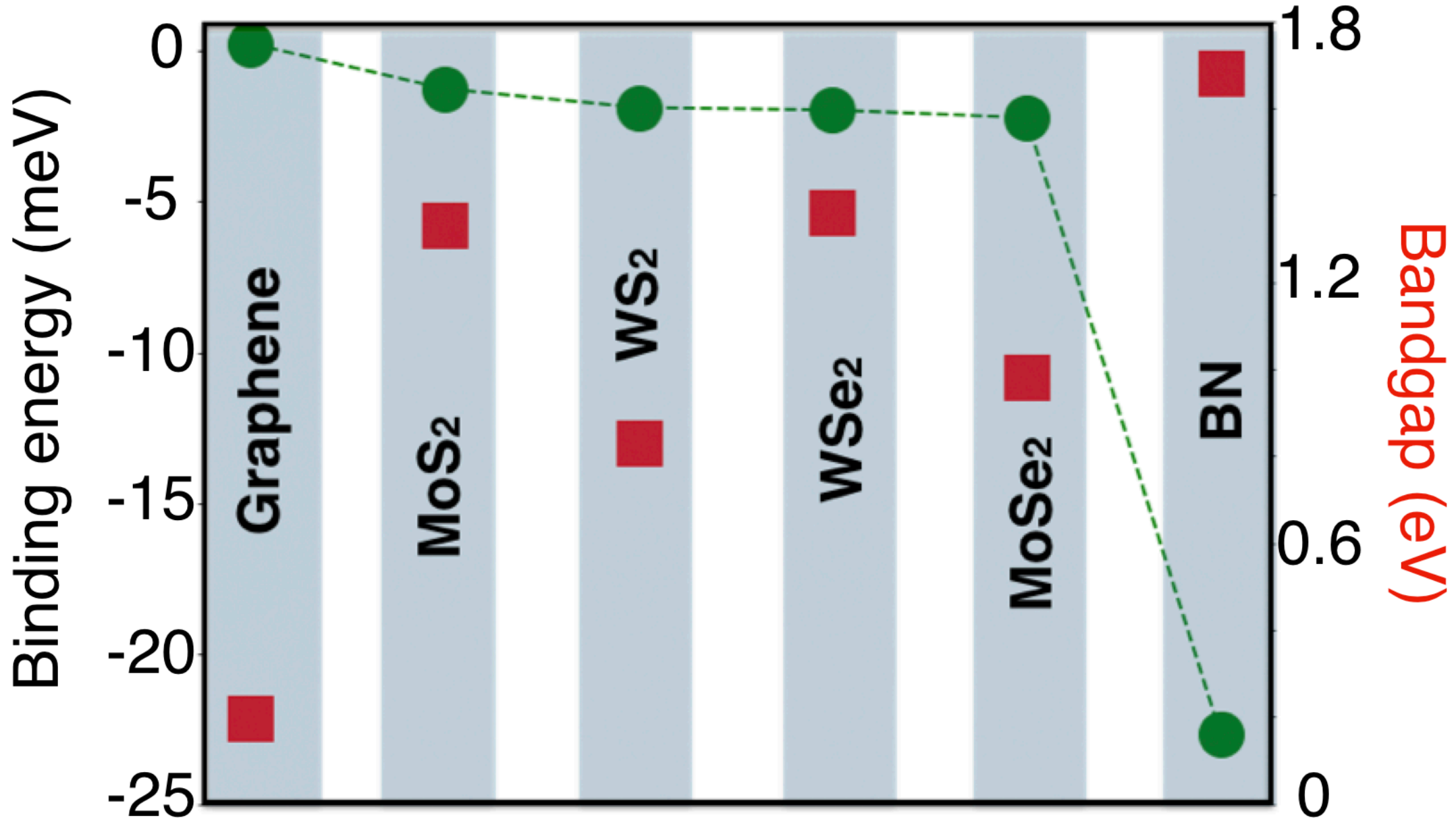

**Figure 2:** We plot binding energy (solid-circles green) and bandgap (solid-square red) for the six *$MAPbI_3ML$ deposited* (*$MoS_2$*, *$MoSe_2$*, *$WS_2$*, *$WSe_2$*), *BN* and *Graphene heterostructures*. The *BN-ML/ $MAPbI_3ML$* is the most stable with a bandgap of ~1.68 eV, which is very close to the optimal gap ~1.4 eV.

(a)

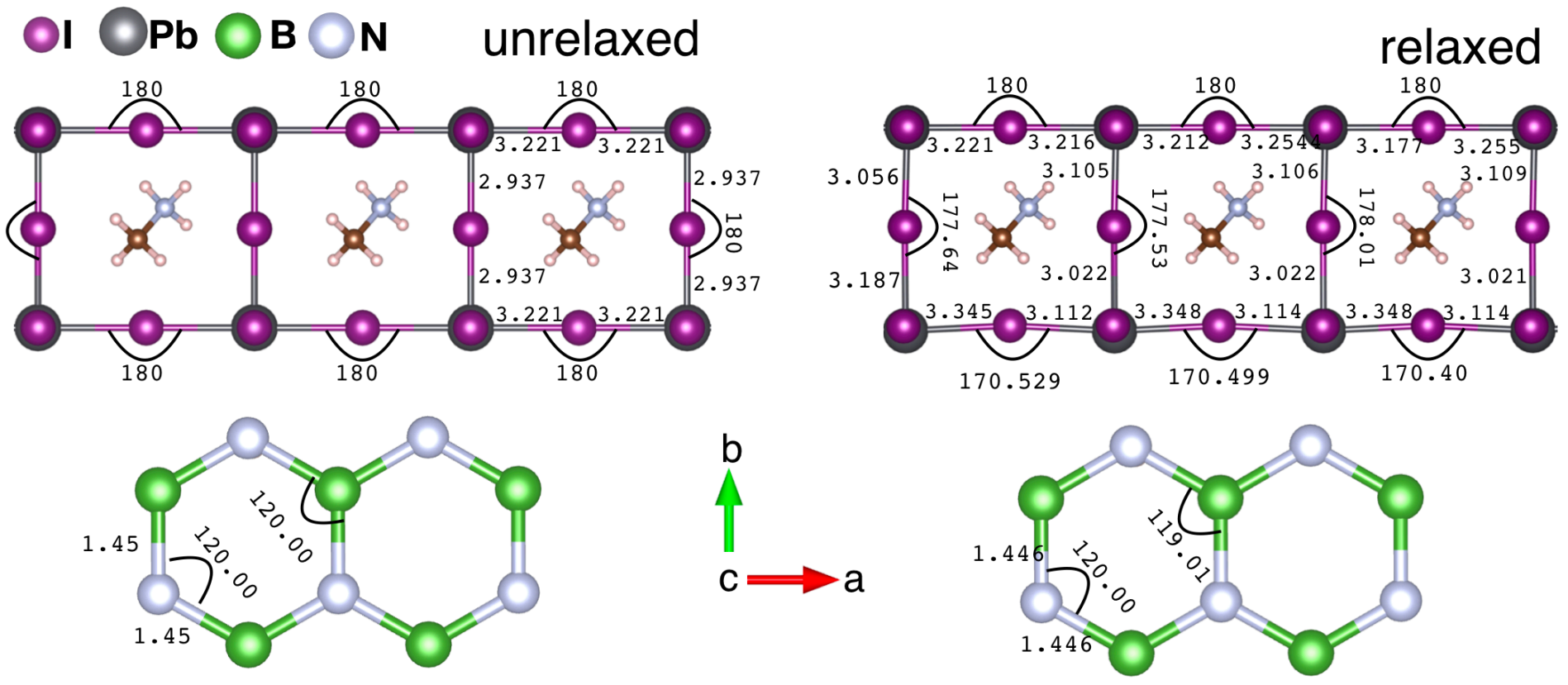

(b)

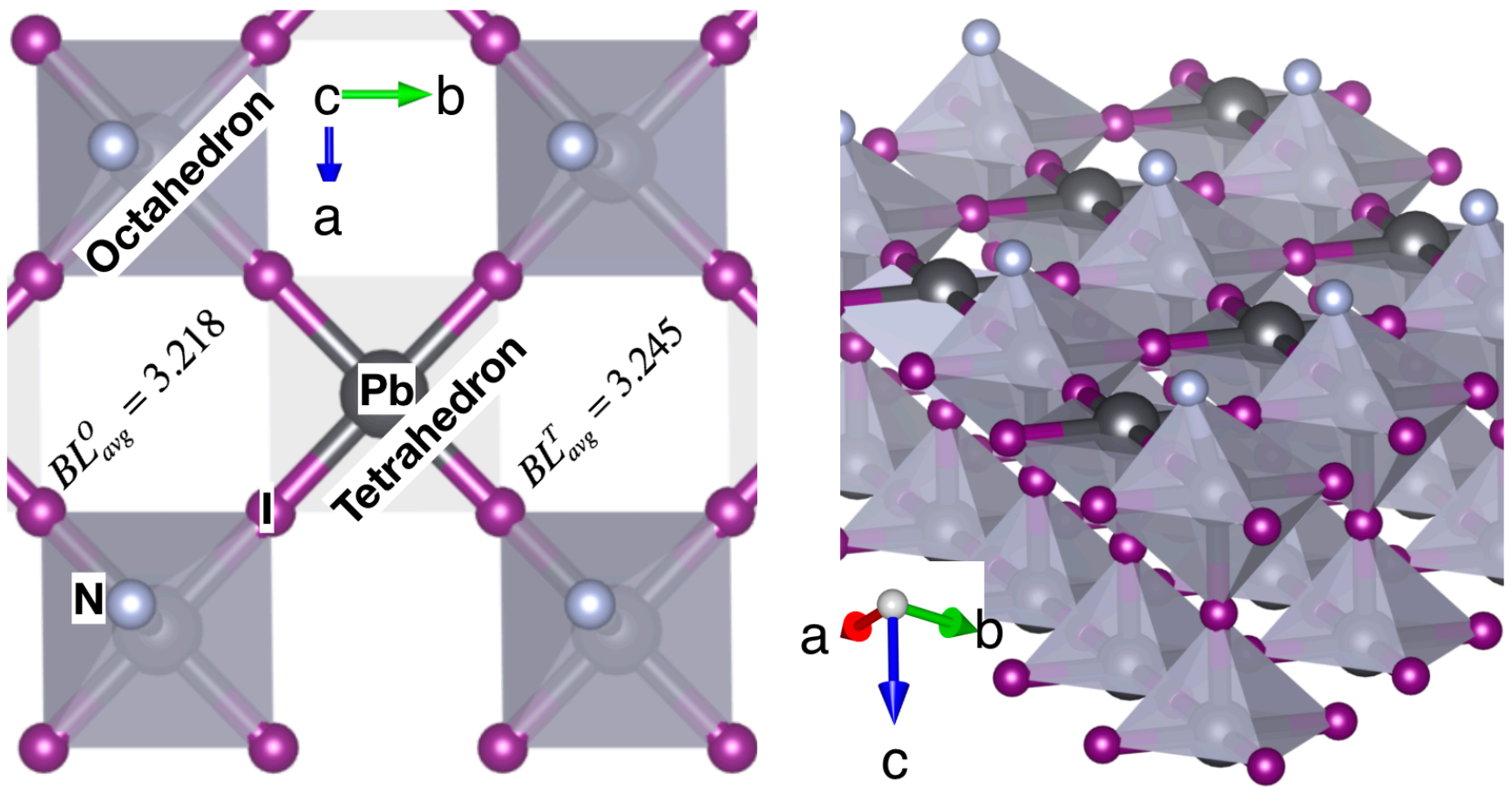

**Figure 3:** (a) The unrelaxed (left-panel) and relaxed (right-panel) structure of *BN-ML/MAPbI₃ML*. The bond-length of *BN-ML* shows small change from 1.45 Å (unrelaxed) to 1.446 Å (relaxed), while only *N-B-N* shows change in bond-angle from 120° (unrelaxed) to 119.01°. (b) The relaxed *MAPbI₃ML* on *BN-ML* is arranged in a maze shape, where every alternate *Pb* at the interface form octahedron with *N*-atom of the *BN-ML*. The average bond-length of octahedron and tetrahedron is 3.218 Å and BL(T) = 3.245 Å, respectively.

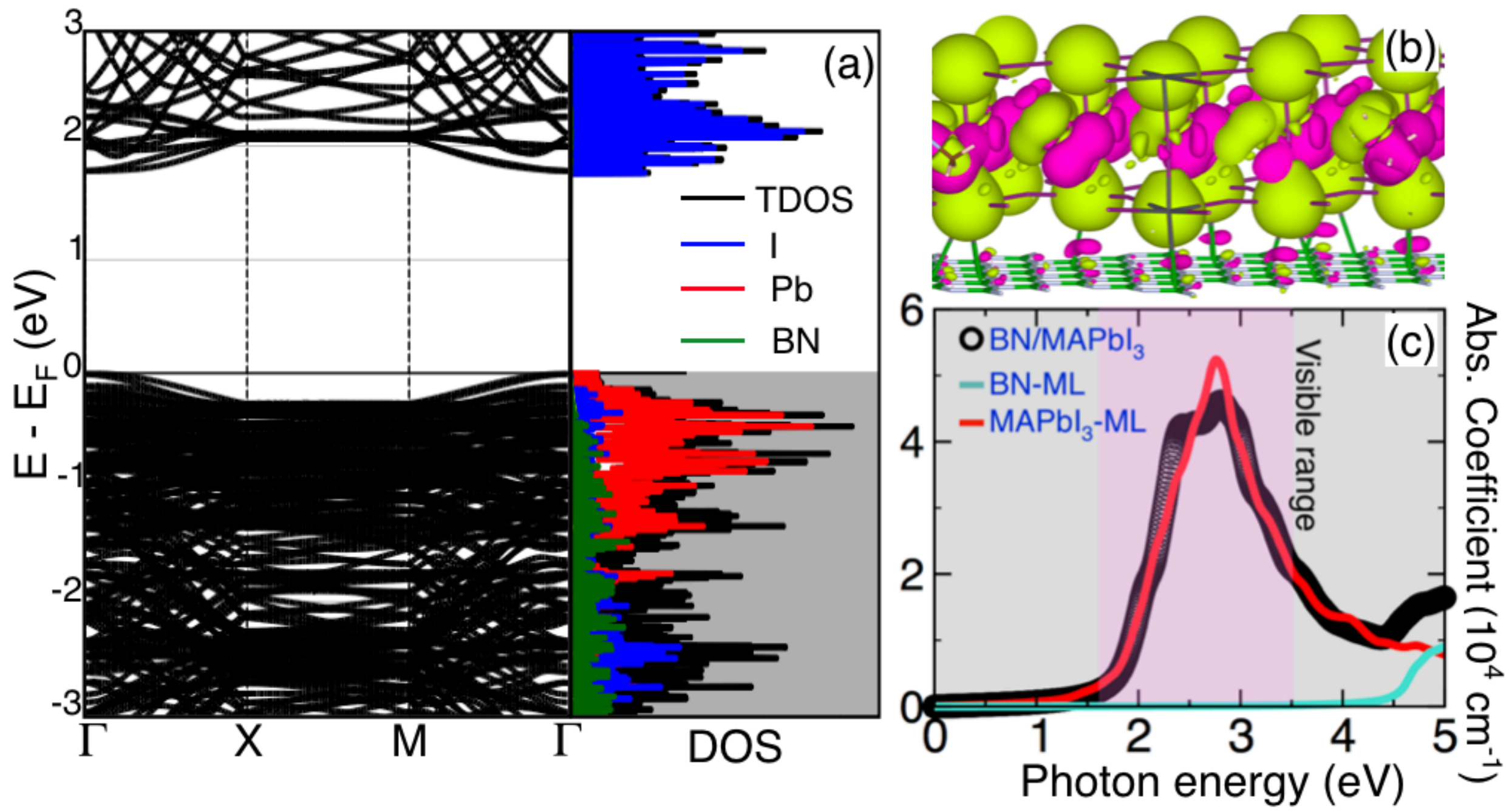

**Figure 4:** (a) The electronic band-structure and density of states (total and projected with specific contributions of *I-5p*, *Pb-6s* and *BN-p*) for *BN-ML/MAPbI₃ML*. The calculated direct bandgap at Γ is 1.68 eV. The significant contribution in band-structure comes from *Pb*-6*s* and *I*-5*p* present near the Fermi level. (b) In charge density difference, distorted *Pb-6s (green) lobe and small contribution from BN-p (red) states show interlayer interaction due to* small charge transfer. (c) The absorption coefficient of *BN-ML/MAPbI₃ML* is in the visible range of light from 1.69 to 3.50 eV.

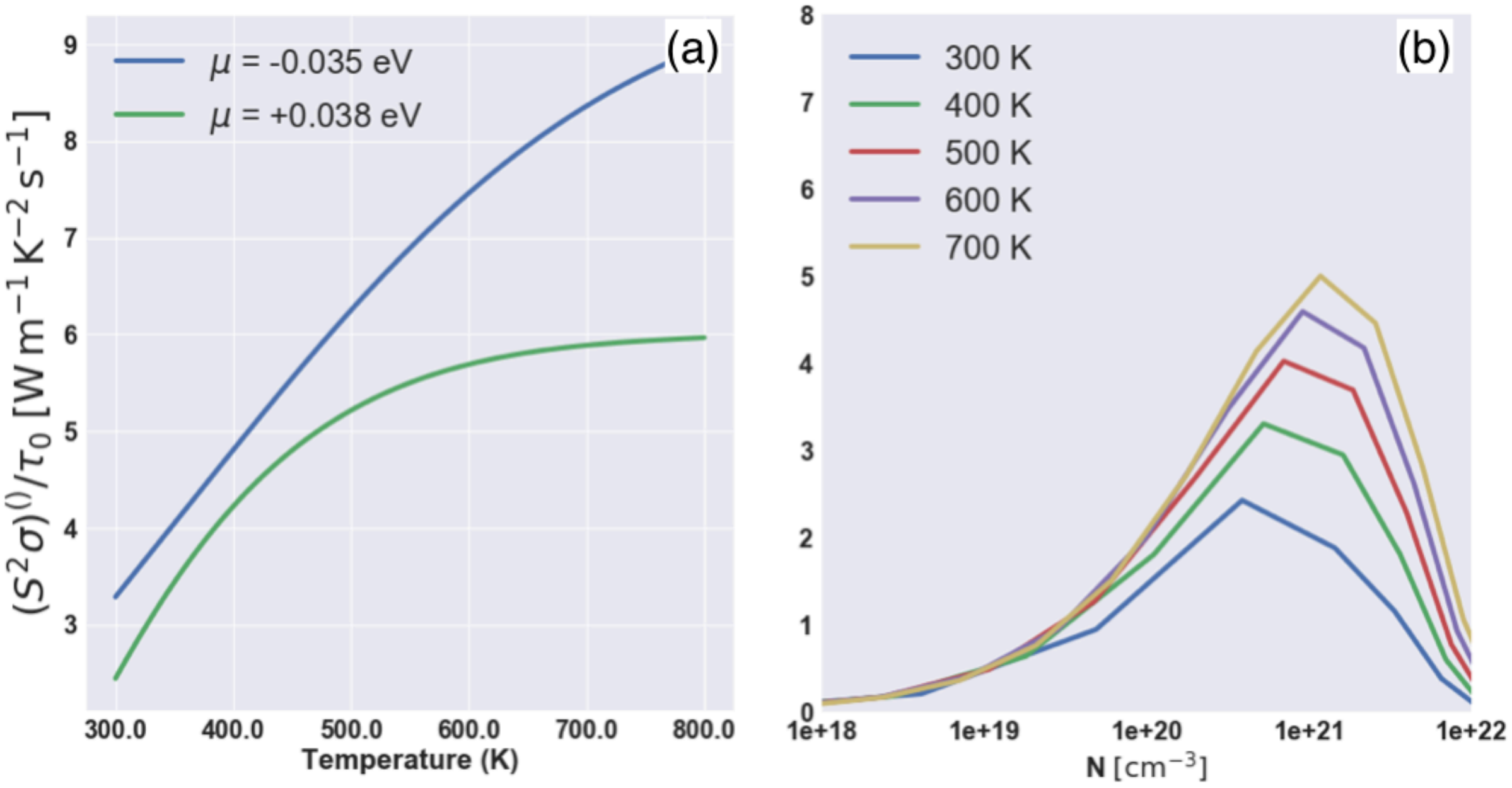

**Figure 5:** The power factor ($= \frac{S^2\sigma}{\tau}$) shown with (a) temperature (K), and (b) number of charge carriers ($cm^{-3}$). Power factor increases with temperature, which is an indicative of the increase in number of charge carriers in conduction band.

Table 1. Table: *Effective mass of BN-ML/MAPbI$_3$ML along X and M direction of Brillouin zone*.

| | VB | | CB | |
|---|---|---|---|---|
| | X | M | X | M |
| $m_{xx}$ | 0.916 | 0.384 | 0.143 | 0.156 |
| $m_{yy}$ | 0.076 | 0.145 | 0.050 | 0.042 |